# Crystallographic Study of U-Th bearing minerals in Tranomaro, Anosy Region- Madagascar

F. E. Sahoa[a,*], N. Rabesiranana[a], Raoelina Andriambololona[a], H. Geckeis[b], C. Marquardt[b], N. Finck[b]

[a]Institut National des Sciences et Techniques Nucléaires (Madagascar-INSTN)
B.P. 4279 - 101 Antananarivo, MADAGASCAR
E-mail: instn@moov.mg

[b]Karlsruhe Institute of Technology (KIT) - Institute of Nuclear Waste and Disposal (INE)
Hermann-von-Helmholtz Platz1
D-76344 Eggenstein-Leopoldshafen, GERMANY
E-mail: horst.geckeis@kit.edu, christian.marquardt@kit.edu, nicolas.finck@kit.edu

As an alternative to conventional fossil fuel, there is a renewed interest in the nuclear fuel to support increasing energy demand. New studies are then undertaken to characterize Madagascar U-Th bearing minerals. This is the case for the urano-thorianite bearing pyroxenites in the south East of Madagascar. In this region, several quarries were abandoned, after being mined by the French Atomic Energy Commission (C.E.A) in the fifties and sixties and are now explored by new mining companies. For this purpose, seven U-Th bearing mineral samples from old abandoned uranium quarries in Tranomaro, Amboasary Sud, Madagascar (46° 28' 0"E, 24° 36' 0"S), have been collected. To determine the mineral microstructure, they were investigated for qualitative and quantitative identification of crystalline compounds using X-ray powder diffraction analytical method (XRD). Results showed that the U and Th compounds, as minor elements, are present in various crystalline structures. This is important to understand their environmental behaviours, in terms of crystallographic dispersion of U-Th minerals and their impacts on human health.

## 1. INTRODUCTION

In 1907, a Malagasy mineralogist, J.B. Rasamoel, discovered euxenite-bearing tertiary lake sediments near Vinaninkarena. French interest in the source of radium lead to intensive search for primary uranium minerals in the adjacent upland plateau [1]. The central island is one of the regions of interest of a Manhattan District Project. In 1943-1944, the Kitsamby district, located in the region, was then explored for potential uranium ore deposit by the project specialists [2]. Moreau has observed uranium - thorium mineralization in Tranomaro [3]. To support the French effort to develop nuclear project, major extracting activities were undertaken by the French Atomic Energy Commission (CEA) during the fifties and sixties. In 1979, using new analytical technology such as gamma spectrometry system, Raoelina Andriambololona et al. have further investigated the radioactivity characteristics of U-Th minerals from the South-East of Madagascar [4,5,6]. Rakotondratsima [7], using electronic microprobe analysis and XRF technique, studied the origin and concentration of uranothorianite bearing pyroxenite content in the region, and has demonstrated the formation of uranothorianite bearing pyroxenites with marble at Belafa and Marosohihy and mineralization without marble at Ambidandrakemba, two billion years ago. Rakotondrazafy used XRD technique and microprobe analysis for dating and determining the formation conditions of hibonite in Madagascar. He has indicated that hibonite mineral has been plainly confined in urano-thorianite bearing skarns formed 545 million years ago [8].

Recently, in order to supply worldwide increasing need for alternative energy, and in view of future shortage of conventional fossil fuel, mining companies are exploring old and new mining areas. This is the case for Tranomaro region, where is a renewed interest for uranothorianite deposits observed. This study is then undertaken to support the risk analysis related to the dispersion of radioactive elements in the environment from old and future mining activities. For this purpose, crystallographic structure of the U-Th compounds in pyroxenite mineral samples is investigated using X-ray powder diffraction analytical technique (XRD).

## 2. METHODS

Radiometric measurement within the study area has already been done by the Pan African Mining Corporations-Madagascar team. Using Exploranium scintillator counters, they have localized high radioactivity spots. Sampling points are then selected according to counting level. Only samples with count rate between 1650 and 3000 counts per second (cps) have been selected. They are usually from abandoned mining quarries and details are shown in Table-1. Most of the samples are crude rocks and one of them is a pure U-Th mineral carefully picked from a pyroxenite sample.

**Table 1.** Mineral samples and their localization.

| Code | Mine Site | Counts (cps) | Longitude | Latitude | Locality |
|---|---|---|---|---|---|
| R01M37 | M37 | 1610 | 46 °30'24''E | 24°34'43''S | Ambindandrakemba |
| R01M47 | M47 | 1650 | 46 °31'47''E | 24°23'36''S | Marosohihy |
| R02M47 | M47 | 2500 | 46 °31'46''E | 24°23'36''S | Marosohihy |
| R03M47 | M47 | 3000 | 46 °31'45''E | 24°23'38''S | Marosohihy |
| R01M52 | M52 | 1720 | 46° 31'36''E | 24° 21'47''S | Belafa |
| R02M52 | M52 | 1725 | 46° 31'48''E | 24° 21'55''S | Belafa |
| Uranothorianite | | 2800 | 46 °30'27''E | 24°34'12''S | Uranium ore |

Mineral samples are grinded and sieved in order to get 10µm to 20µm size fine and homogeneous powder. For low diffraction background, they are packed into a sample holder made of silicon single crystal and smeared uniformly onto a glass slide, assuring a flat upper surface pack into a sample container. Care is taken to create a flat upper surface and to achieve a random distribution of the crystallite lattice orientations.

The XRD analysis is carried out using a D8 Advance diffractometer (Bruker) equipped with a copper anode (40 kV, 40 mA). The sample is illuminated by the X-ray ($\lambda = 1.5406$ Å) and the diffracted beam is detected with an energy dispersive detector (Sol-XE), controlled by the XRD Commander software (Bruker). The diffractograms are collected for $2\theta$ varying from 5 to 80°, with steps of 0.01° and 2 seconds counting time per step with the sample spinning at 15 rotations per minute. The diffratograms displaying the diffracted intensity as a function of the diffraction angle are processed using the Eva software (Bruker). Since each pure crystalline phase is characterized by a specific diffraction pattern, the identification of the phases present in the sample is achieved by comparison with the JCPDS database. A phase is present in the sample when the three most intense peaks from the reference data are present on the experimental diffractogram. If more peaks are present, higher confidence is placed in the presence of the phase. To improve the accuracy, only phases which are likely to be found in the studied samples according to complementary results are selected for identification. From literature reviews and crystallographic handbooks, table-2 shows such (U,Th) compounds and major minerals in pyroxenite [7,8].

**Table 2**. Minerals expected to be found in the studied (U-Th) rocks

| Expected elements | Chemical formula |
|---|---|
| Uranothorianite | $(Th,U)O_2$ |
| Diopside pyroxenite | $CaMg(SiO_3)_2$ |
| Calcite | $CaCO_3$ |
| Corundum | $CaMg(SiO_3)_2$ |
| Spinel | $FeAl_2O_4, MgCr_2O_4$ |
| Apatite | $Ca_5(PO_4)_3(Cl,F,OH)$ |
| Phlogopites | $KMg3(AlSi_3)O_{10}(F,OH)_2$ |
| Hibonite | $Ca\ Al_{12}O_{19}$ |

## 3. RESULTS AND DISCUSSIONS

Two typical diffractograms for the samples uranothorianite and R01M37 are shown in Figures 1 and 2, respectively. The first one is a simple spectrum from pure (U,Th) minerals. The second one is a complex spectrum from a natural (U,Th) ore.

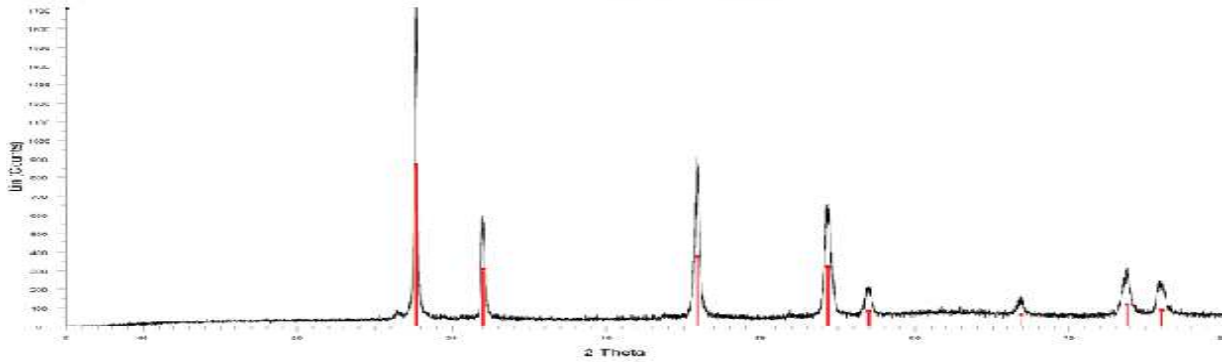

**Figure 1.** Pure XRD spectrum from uranothorianite minerals

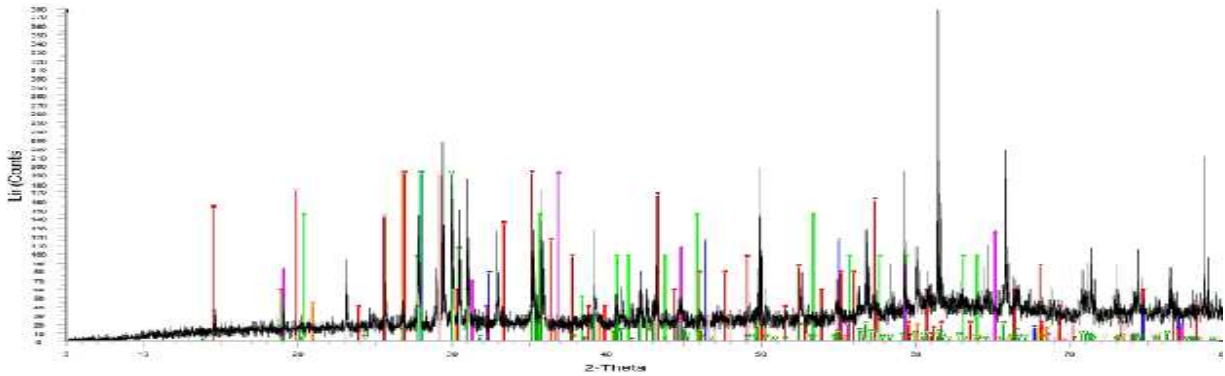

**Figure 2.** Complex XRD spectrum from R01M37 sample

For natural ores, common major and minor mineral phases such as corundum, spinel, diopside, soddyite, calcium silicate and calcite are usually present in the samples. Interestingly, (U-Th) compounds are present in multiple crystalline phases and systems, as shown in Table-3.

**Table 3.** Crystalline systems of minerals detected by XRD analysis

| Sample code | Identified compounds | Chemical formula | Crystalline systems |
|---|---|---|---|
| Uranothorianite | Thorianite | $ThO_2$ | cubic |
| R01M37 | Thorium uranium oxide | $Th_{0.5}U_{0.5}O_{2.04}$ | cubic |
|  | Iron uranium oxide | $Fe_2UO_6$ | hexagonal |
| R01M47 | Uranium oxide | $UO_3$ | monoclinic |
| R02M47 | Uranium oxide | $UO_3$ | monoclinic |
|  | Uranium oxide | $U_2O_5$ | orthorombic |
| R03M47 | Uranium oxide | $UO_3$ | monoclinic |
|  | Uranium oxide phosphate | $U_2P_2O_{10}$ | complex |
| R01M52 | Uranium oxide | $U_3O_7$ | tetragonal |
|  | Uraninite | $UO_2$ | orthorombic |
|  | Thorianite | $ThO_2$ | cubic |
| R02M52 | Thorianite | $ThO_2$ | cubic |
|  | Uranium oxide | $UO_3$ | complex |

The uranothorianite sample contains thorianite as $ThO_2$ in cubic system. R01M37 contains $Th_{0.5}U_{0.5}O_{2.04}$ in cubic system, which can be considered as a mixture of $UO_2$ and $ThO_2$, and $Fe_2UO_6$ in hexagonal system, which can be considered as a mixture of $Fe_2O_3$ and $UO_3$. R03M47 contains $UO_3$ in monoclinic system and $U_2P_2O_{10}$ as a complex system, which can be considered as mixture of $U_2O_5$ and $P_2O_5$. R02M47 contains $UO_3$ in monoclinic system and $U_2O_5$ in orthorhombic system. R01M47 contains only $UO_3$ in monoclinic phase.

R01M52 contains thorianite as $ThO_2$ in cubic system, uranium oxide as $U_3O_7$ in tetragonal system and uraninite as $UO_2$ in orthorhombic system. R02M52 contains thorianite as $ThO_2$ in cubic system and uranium oxide as $UO_3$ in complex phase.

The presence of both thorianite and uraninite is expected. According to Frondel, uraninite is commonly present in pegmatites [9]. Moreau explained that previously, thorium and uranium were rather concentrated in the granitic and charnocktic zones. Precambrian granitization is followed by the formation of the main pegmatitic areas in Madagascar with Th-U niobotantalates, uraninite and beryl. In the same time, a large uranium and thorium provinces with uranothorianite deposits appear within the calcomagnesian series of the southern area of Madagascar [3].

Thorianite is an isometric $ThO_2$ crystal (found in R01M37, R01M52, R02M52 and uranothorianite samples) in the hexoctahedral class $\left(\frac{4}{m} \overline{3} \frac{2}{m}\right)$ sometimes modified to {111} or {113}[9]. The uranothorianite diffractogram (Fig.1) particularly shows the dominance of $ThO_2$ in uranothorianite sample. This demonstrates that thorianite is the main component of the Tranomaro U-Th system.

Various definitions of uraninite are proposed by specialists. Following the Dana classification, uraninite is defined as an oxide form of U and Th and has an isometric crystalline structure. According to Wasserstein [10,11], who proposed a classification based on the radiogenic lead content and the redox conditions, α-uraninite corresponds to $UO_2$ (found in R01M37 and R01M52 samples), β-uraninite to $U_3O_7$ (found in R01M52 sample) and τ-uraninite to $U_4O_9$. It is also known that $UO_2$ is transformed into $U_3O_8$ and $UO_3$ (found in R01M37, R01M47, R02M47, R03M47 and R02M52 samples) under oxidizing conditions. Ebelmen reported also that $U_2O_5$ (found in R02M47 and R03M47 samples) is an essential constituent of uraninite [12].

Uraninite resembles closely to thorianite. This is because U and Th are isometric and both present in minerals in different concentrations (88.15 % of U in uraninite and 70% of Th in thorianite). Some authors have proposed a solid solution model to explain the possibility of U-Th exchange between the uraninite and thorianite systems [13,14,15], but their oxidation behaviours differ from each another. On the one hand, this results in a simple crystal system for thorianite: it only exists in a cubic system. On the other hand, the uranium component of the Tranomaro uranothorianite samples are oxide based and is a mixture of complex oxidations states and crystalline systems. This is in agreement with published results, showing that uraninite is a complex uranium oxide compounds [10,11,12,13,14,15].

## 4. CONCLUSION

The study of mineral samples collected from Tranomaro area using X-ray diffraction gives a good understanding of the crystalline structure of the (U-Th) minerals. From seven collected and analysed samples, the thorium oxide is identified in a simple crystalline system. On the opposite, the uranium oxides show multiple oxidation states and crystalline systems, including α-uraninites and β-uraninite. Due to significant presence of uraninite and thorianite which are radioactive minerals, populations living in the Tranomaro area may be exposed to increased environmental radiation. As each crystalline system is more or less subject to dissolution or dispersion in the environment, further investigation must be done to investigate the details of their transport behaviours in the environment.

**Acknowledgments**

This paper was completed within the framework of a cooperation between the Institut National des Sciences et Techniques Nucléaires (Madagascar-INSTN) and the Institute of Nuclear Waste and Disposal (INE). The German Academic Exchange Service (Deutscher Akademischer Austausch Dienst: DAAD) financed the fellowship. The Pan African Mining-Madagascar (PAM) facilitated the access to the Tranomaro sites. Technical supports from Madagascar-INSTN and INE are gratefully acknowledged.